\documentclass[twocolumn,showpacs,preprintnumbers,amsmath,amssymb]{revtex4}

\usepackage{graphicx}
\usepackage{dcolumn}
\usepackage{bm}
\usepackage{natbib}
\begin{document}


\title{Analysis of noise-induced bistability in Michaelis Menten single-step enzymatic cycle} 

\author{Daniel Remondini}
\email{daniel.remondini@unibo.it}
\author{Enrico Giampieri}
\author{Armando Bazzani}
\author{Gastone Castellani}
\affiliation{Physics Dept. of Bologna University and INFN Bologna}%
\author{Amos Maritan}
\affiliation{Physics Dept. of Padova University} %

\date{\today}

\begin{abstract}
In this paper we study noise-induced bistability in a specific circuit with many biological implications, namely a single-step enzymatic cycle described by Michaelis Menten equations with quasi-steady state assumption.
We study the system both with a Master Equation formalism, and with the Fokker-Planck continuous approximation, characterizing the conditions in which the continuous approach is a good approximation of the exact discrete model. An analysis of the stationary distribution in both cases shows that bimodality can not occur in such a system.
We discuss which additional requirements can generate stochastic bimodality, by coupling the system with a chemical reaction involving enzyme production and turnover.
This extended system shows a bistable behaviour only in specific parameter windows depending on the number of molecules involved, providing hints about which should be a feasible system size in order that such a phenomenon could be exploited in real biological systems.
\end{abstract}


\maketitle

\section{Introduction}

Many biological phenomena (e.g. memory induction, chromatin remodeling, cell-fate determination) are receiving a great deal of attention in recent years, due to an increasing interest in the description of their complex behaviour by means of basic biochemical circuitry \cite{Aguda11,Lahav11,VanOud07}.
The  signal transduction machinery is mainly based on enzymatic reactions, whose average kinetics can be described within the framework initally proposed by Michaelis and Menten (MM). The steady state approximation of MM model accounts for the majority of known enzymatic reactions, and can be adjusted for the description of regulatory properties such as cooperativity, allostericity and activation/inhibition \cite{Xie2006}. The MM equations are still valid at small molecule numbers (as it frequently happens in real cells) if the microscopic interpretation is changed correspondingly \cite{QianElson2002,English2006,Gill02} but the discrete stochastic aspects become predominant, and a deterministic or a stochastic continuous model can not describe the system in sufficient detail \cite{VanOud2008,Gupta10,Maheshri10}.

A large class of enzymatic reactions controls the reversible addition and removal of phosphoric groups, phosphorylation/dephosphorylation reactions catalyzed by kinases and phosphatases respectively. The phospho/dephosphorylation cycle (PdPC) is thus a post-translational substrate modification that is central for the regulation of several biological processes \cite{Krebs1958,Samoilov2005}.

How these processes can show a bistable behaviour \cite{Ortega2006,Cast09} in the presence of fluctuations \cite{Bergetal2000}, reflected by a bimodal stationary distribution of protein number/concentration, is a crucial question for their modeling.

The deterministic version of a single PdPC is not bistable in general, but it is hypothesized that external noise can trigger such a behavour \cite{Samoilov2005}.
We study this cycle by a Chemical Master Equation approach, and also (in the limit of large molecule number) the related Fokker-Planck equation.
A closed form for the stationary distribution of the system is obtained for both approaches: we show that the system can not have a bimodal stationary distribution due to intrinsic fluctuations only, but an additional external noise obtained by a plausible biological mechanism  (i.e. the coupling of the system with an enzyme production/activation reaction) can produce such feature.
We define analytically the conditions in which bimodality occurs, as a function of the reaction parameters (kinetic constants) and system size (number of enzyme and substrate molecules), and we verify our results by numerical simulations with a Gillespie algorithm.

\section{The model}

The PdPC (also referred to as the futile cycle) is composed by one phosphorylation and one dephosphorylation reaction, catalyzed respectively by enzymes $E_1$ and $E_2$:
\begin{eqnarray}
E_1 + A \rightleftharpoons E_1A \rightarrow A^P + E_1 \nonumber \\
E_2 + A^P\rightleftharpoons E_2A^P  \rightarrow A + E_2
\label{schema}
\end{eqnarray}
The deterministic dynamics of this cycle can be described via the MM formalism. Assuming a steady-state approximation for both enzymatic reactions $\dot{A}$ and $\dot{A^P}$, we obtain the following equations:
$$
\dot{A^P} = v_1 - v_2 \qquad \dot{A}= v_2 - v_1
$$
where
\begin{eqnarray}
v_1&=&K_{C1}\cdot E_{1}\frac{A}{K_{M1} + A} = V_{M1}\frac{A}{K_{M1} + A}\nonumber\\
v_2&=&K_{C2}\cdot E_{2}\frac{A^P}{K_{M2} + A^P} = V_{M2}\frac{A^P}{K_{M2} + A^P}
\label{MM}
\end{eqnarray}
Imposing the conservation of the total substrate concentration, let $x$ be the $A$ molecule concentration, we obtain:
\begin{equation}
\dot{x} = V_{M2}\frac{1-x}{K_{M2} + 1-x} - V_{M1}\frac{x}{K_{M1} + x},
\label{DetEq}
\end{equation}
that can be easily shown to have only one solution inside the substrate domain (see \cite{Samoilov2005}).

\subsection{The CME approach}

Starting from the previous equations, a Chemical Master Equation (CME) approach \cite{VanKampen} is introduced to account for intrinsic noise ($p_n$ is the $A$-molecule distribution function over the possible states $n\in [0:N]$, $D_+f(n)=f(n+1)-f(n)$):
\begin{equation}
\dot{p_n} = D_+ J; \quad J=r_n p_n- g_{n-1} p_{n-1}
\label{CME_fc}
\end{equation}
where
$$
r_n = V_{M1}' \frac{n}{K_{M1}' + n}  \qquad g_n =  V_{M2}' \frac{N-n}{K_{M2}' + N-n}
$$
$N$ is the total number of the substrate molecules, $n$ is the the number of $A$ molecules and the MM constants have been accordingly scaled: $K_{M}'=N\cdot K_{M}$ and $V_M'=N\cdot V_M$. 
In the hypothesis of fast relaxation times, the stationary solution of eq. (\ref{CME_fc}) describes the statistical properties of the reaction in Fig. \ref{schema}.
The stationary distribution $p_n^s$ is derived by imposing $\dot{p_n}(t)=0$; 
excluding the existence of a constant current in the system, we get the condition
$$
\frac{p^s_{n}}{p^s_{n-1}}=\frac{g_{n-1}}{r_n}\quad \Rightarrow \quad D_+ \ln p^s(n)=\ln \frac{g_{n}}{r_{n+1}}
$$
If we define a potential $V(n)$, such that $D_+V(n)=-\ln ({g_{n}}/{r_{n+1}})$, the stationary solution has the Boltzmann form
\begin{equation}
p^s_{n}=F\cdot e^{-V(n)},
\label{boltz}
\end{equation}
where $F$ is a normalizing constant. According to (\ref{boltz}), the maxima and minima of the distribution are obtained by imposing $ D_+V(n)=0$, extending the $n$ domain to the set of real numbers. This leads to ${g_{n}}/{r_{n+1}}=1$, similar to (\ref{DetEq}) and with an unique solution inside the $[0:N]$ domain. This result is also confirmed by Gillespie simulations of the dynamical process.

\subsection{Fokker-Planck approximation}

CME dynamics can be more easily studied by considering a continuous approximation by means of a Fokker-Planck (FP) equation, with $N$ is sufficiently large to get the natural boundary conditions $p^s(1)=p^s(0)\simeq 0$. Expanding in power series the $D_+$ operator up to second order, from eq. (\ref{CME_fc}) we get:
\begin{equation}
\frac{\partial p}{\partial t}(x,t) =
-\frac{\partial}{\partial x} C(x) p(x,t)+\frac{1}{2}\frac{\partial^2}{\partial x^2} D(x) p(x,t)
\label{fp_eq}
\end{equation}
where the drift and diffusion coefficients are defined as $ C(x) = g(x)-r(x)$, $D(x) = \left[r(x)+g(x)\right]/N $, $x=n/N$ ($0\leq x \leq 1$), $p_n = p(x)/N$, and
\begin{align*}
g(x) & = \frac{V_{M2}\cdot(1-x)}{K_{M2}+1-x};\\
r(x) & = \frac{V_{M1}\cdot x}{K_{M1}+x}
\end{align*}
up to an error $O(N^{-1})$.
We remark that the diffusion coefficient becomes negligible for large $N$ with respect to the drift term: the fluctuations scale as $1/\sqrt{N}$ according to the law of large numbers.
The stationary solution $p^s(x)$ can be written explicitly:
\begin{equation*}
p^s(x)  =  \frac{F}{D(x)}\exp\left (2 \int^{x}\frac{C(y)}{D(y)}dy\right ) 
\end{equation*}
that for the symmetric case $V_{M1}=V_{M2}$, $K_{M1}=K_{M2}=K_M$ reduces to
\begin{equation}
p^s(x)  =F \left( K_M+1-x\right) \left( K_M+x\right) \left( K_M+2x-2x^2\right)^{K_M-1}
\label{FP_staz}
\end{equation}
where $F$ is a normalizing constant. 

\begin{figure}[htbp]
\includegraphics[width=0.4\textwidth]{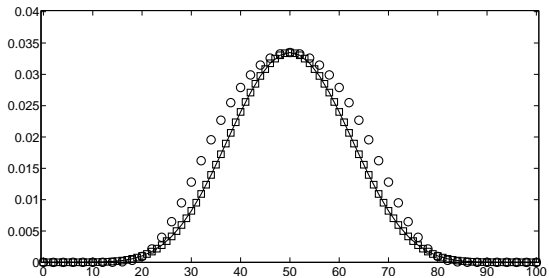}
\caption{
Solutions for the futile cycle: $N=100$, $V_M = 1$, $K_M = 0.1$, $E_1 = E_2$. Squares: CME stationary solution; line: FP exact solution; circles: FP approximate solution (\ref{gauss_app}).
} \label{CME_FP}
\end{figure}

The FP stationary solution is a good approximation of CME when the drift coefficient is of the same order of the diffusion coefficient (i.e. $C(x)\simeq O(1/N)$, see Appendix).
Therefore, we can apply the FP approximation nearby the critical points $x_\ast$ where $C(x)\simeq 0$. 
By using a Gaussian approximation, in the neighborhood of a critical point $x_\ast$ we get the distribution ($C'(x)=dC/dx$)
\begin{equation}
p^s(x)\propto \exp\left ( -\frac{|C'(x_\ast)|}{D(x_\ast)}\frac{(x-x_\ast)^2}{2} \right )
\label{gauss_app}
\end{equation}
In the symmetric case, for generic enzyme concentration, we explicitly compute
$$
C(x)=K_C\left ( E_1\frac{1-x}{K_M+1-x}-E_2\frac{x}{K_M+x}\right )
$$
so that
\begin{equation}
\begin{split}
\frac{C'(x)}{D(x)}=-K_M N\left ( \frac{E_1}{(K_M+1-x)^2}+\frac{E_2}{(K_M+x)^2}\right ) \\ 
\left ( \frac{E_1(1-x)}{K_M+1-x}+\frac{E_2x}{K_M+x}\right )^{-1}
\end{split}
\end{equation}
The variance $D(x_\ast)/|C'(x_\ast)|$ scales as $(K_M N)^{-1}$ and, in general, is weakly dependent on the enzyme concentration. 
Therefore, in the continuous limit $N\gg 1$, we get a finite variance only for $K_M\ll 1$. 
In Fig. \ref{CME_FP} we compare the exact CME stationary distribution, the exact solution of the FP equation and the approximation of Eq. (\ref{gauss_app}) in the symmetric case, where the condition $C(x)\ll 1$ is satisfied. 
In the FP equation (\ref{fp_eq}) two minima may appear in the physical domain of $x$ for specific values of $K_M$ only when $N$ is sufficiently low, but these critical points have no meaning being due to a bad approximation of the CME. Thus intrinsic noise cannot induce stochastic bifurcation in system (\ref{schema}). 

\subsection{Bimodality induced by enzyme noise}

Relaxing the assumption of fixed enzyme concentration, we characterize the effect of enzyme fluctuations on the substrate stationary distribution.
In the symmetric case $K'_{M1}=K'_{M2}=K'_M$and $K'_{C1}=K'_{C2}$,  the equilibrium points of the average equation (\ref{MM}) corresponding to maxima of $p^s_n$ can be calculated explicitly  as a function of the ratio between enzymes $\gamma$:  
\begin{equation}
\gamma =\frac{E_{1}}{E_{2}}=\frac{N-n+1}{n}\frac{K'_M+n}{K'_M+N-n+1}
\label{dist_eq}
\end{equation}
If one introduces the variable $u$:
\begin{equation}
u=\frac{n}{N}-\frac{N+1}{2N}=\frac{n}{N}-a\qquad u\in \left[ -a,+a+\frac{1}{N}\right]
\label{u_def}
\end{equation}
where $a\simeq 1/2$ for $N\gg 1$, the condition (\ref{dist_eq}) reads
$$
\gamma = \frac{a-u}{a+u}\cdot\frac{K_M+a+u}{K_M+a-u},
$$
where $K_M = K'_M/N$.
Assuming that $K_M\ll 1$, so that the critical point is quite sensitive to the enzyme concentration, and performing a perturbative approach over $K_M$, the previous equation can be rewritten as
\begin{equation}
\left (\frac{1}{a+u}-\frac{1}{a-u}\right )=\frac{\gamma-1}{K_M}
\label{dist_eqs}
\end{equation}
When $\gamma=1$ (i.e. $E_1 = E_2$) we have the trivial solution $u=0$ (an unique maximum with $n = (N+1)/2$, $x=1/2$), whereas for $\gamma-1>0$ (resp. $<0$) $u$ shifts towards $-a$ (resp. $a$).\\
Supposing that enzyme concentration can fluctuate around the average value, given $\xi=(\gamma-1)/K_M$ and $p(\xi)$ the corresponding probability distribution, we have
\begin{equation}
p(u)du=p(\xi(u))\left |\frac{d\xi}{du}\right | du=p(\xi(u))\left (\frac{1}{(a+u)^2}+\frac{1}{(a-u)^2}\right )du
\label{prob_dist}
\end{equation}

Under the hypotheses that $\xi$ fluctuates around zero and $p(\xi)$ tends sufficiently fast to zero at the boundaries (natural boundary condition), we study the conditions for bimodality of $p(u)$. The critical points of $p(u)$ must satisfy
\begin{equation}
\frac{dp(u)}{du} = \frac{d^2\xi}{du^2}p(\xi(u))+\left (\frac{d\xi}{du}\right )^2\frac{dp}{d\xi} = 0
\label{crit_eq}
\end{equation}
If we approximate $p(\xi)$ with a Gaussian distribution (justified for a sufficiently large enzyme molecule number and $K$ sufficiently small, see Fig. \ref{HistChi}) so that
$$
p(u)=\left(\frac{1}{(a+u)^2}+\frac{1}{(a-u)^2}\right) e^{-\frac{1}{2\sigma_\xi^2}\left( \frac{1}{a+u}-\frac{1}{a-u} \right)^2}
$$
the equation (\ref{crit_eq}) reads
\begin{equation}
\begin{split}
\frac{1}{\sigma_\xi^2}\left (\frac{1}{(a+u)^2}+\frac{1}{(a-u)^2}\right )^2\left (\frac{1}{a+u}-\frac{1}{a-u}\right )\\
-2\left (\frac{1}{(a+u)^3}-\frac{1}{(a-u)^3}\right )=0
\end{split}
\label{crit_eq2}
\end{equation}
If we exclude the symmetric solution $u=0$, we get the following condition for bimodality (recalling that $\sigma_\xi=\sigma_\gamma/K_M$)
\begin{equation}
\sigma_\xi^2> \frac{2}{3a^2}\qquad \Rightarrow \qquad \sigma_\gamma^2 > \frac{2K_M^2}{3a^2}
\label{crit_cond}
\end{equation}
The condition (\ref{crit_cond}) can be realized if the enzymatic concentrations fluctuate largely enough around the mean, but for the system (\ref{schema}) this fluctuation cannot be achieved by means of intrinsic noise only.

\section{Stochastic hypersensitivity induces bistability}

\begin{figure}[!ht]
\includegraphics[width=0.5\textwidth]{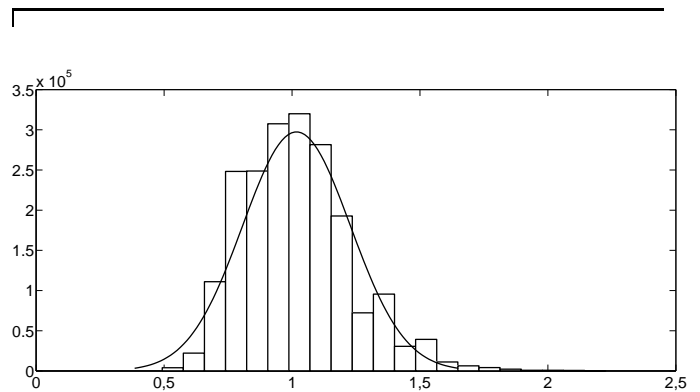}
\caption{Gaussian approximation of $p(\gamma)$ ($\gamma = E_1/E_2$) for $E_1+E_2=100$. Bars: empirical distribution of $\gamma$; continuous line: gaussian distribution with same mean and variance.}
\label{HistChi}
\end{figure}

As we have shown in previous sections, intrinsic noise cannot induce bimodality in PdPC, but large enzyme fluctuations can produce this effect for $K_M$ sufficiently small. This can be achieved by coupling the initial system (\ref{schema}) with a further reaction involving enzyme activation (e.g. by a second messenger like Calcium) or localization  (e.g. inside a delimited region like an organelle). 
Then, in the hypothesis that the activator is in low abundance (or that the reaction region is small) a situation of competition between the two enzymes for the reaction is obtained. 
Under suitable kinetic parameters, the extra fluctuation may lead the cyclic reaction alternately towards one of the two products, producing a bimodal distribution function for the substrates. 
Defining as $E_{1}^{*}$ and $E_{2}^{*}$ the inactive (external to the reaction region) enzyme concentrations, and $E_{A}$ the concentration of activating molecules, the full kinetic reaction scheme becomes:
\begin{eqnarray*}
E_A + E_{1}^{*} \rightleftharpoons E_1 \\
E_A + E_{2}^{*} \rightleftharpoons E_2 \\
E_1 + A \rightleftharpoons E_1A \rightarrow A^P + E_1 \\
E_2 + A^P\rightleftharpoons E_2A^P  \rightarrow A + E_2
\label{A_enz}
\end{eqnarray*}
These equations are identical to those of the enzymatic cycle in (\ref{schema}) as long as there is a sufficiently large number of enzyme molecules, so that the fluctuations in enzyme concentration become negligible.
A simplified version of the enzyme competition can be obtained by considering a direct interchange between the two active enzymes:
\begin{eqnarray*}
E_{1} \rightleftharpoons E_{2} \qquad (E_{2} = E_T-E_{1})
\end{eqnarray*}
with $E_T$ total enzyme concentration. The resulting FP equation of the system is two-dimensional, and the stationary distribution can be expressed as a function of enzyme ratio $\gamma$ and substrate concentration $x$: remembering that $\xi = (\gamma-1)/K_M$ we have $p_s(x,\xi) = \Pi(x|\xi)p_s(\xi)$, in which $\Pi(x|\xi)$ is the conditional distribution of substrate given the enzyme concentration.
Let us consider the case when the PdPC kinetics is much faster than the enzyme (achieved by choosing the $K_C$ parameters in equations (\ref{MM}) sufficiently large) so that one can apply an adiabatic approach to the global system evolution, approximating $\Pi(x|\xi)$ with the stationary solution $p_s(x)$ as in (\ref{FP_staz}). 
The substrate distribution $\overline{p}_s(x)$ is obtained as the marginal distribution of $p_s(x,\xi)$: 
\begin{equation}
p_s(x,\xi) = \int p_s(x|\xi)p_s(\xi)d\xi = \int p_s(x|x_\ast)p_s(x_\ast)dx_\ast,
\label{p_s_s}
\end{equation}
where we use the univocal relationship between the stationary distribution of the enzyme ratio and the equilibrium concentration of the substrate.
Applying the calculation performed previously, we have shown that $p_s(x_\ast)$ is a bimodal distribution if the $E_1/E_2$ distribution can be well approximated by a Gaussian-like function, and if the condition (\ref{crit_cond}) is satisfied.
In Fig. \ref{HistChi} we show that the Gaussian approximation for the enzyme ratio is valid even for a moderate number of enzymes.

For the simplified model in the symmetric case we get a modified version of the previous MM equation (with the steady-state approximation and  $E_1+E_2=1$)
$$
\dot x={{K_{C}\,\left(1-E\right)\,\left(1-x\right)}\over{K_{M}+1-x}}-{{K_{C}\,E\,x}\over{K_{M}+x}}
$$
and we have an explicit dependence of the critical point $x_\ast$ on enzyme concentration $E$:


\begin{widetext}
\begin{equation} 
x_\ast(E)= \frac{2E+K_M-1-\sqrt{1 +2K_M -4E +K_M^2 -8EK_M +4E^2 +8K_ME^2}}{2\left(2E-1\right)}
\label{equilibrioES}
\end{equation}
\end{widetext}

\begin{figure}
\includegraphics[width=0.4\textwidth]{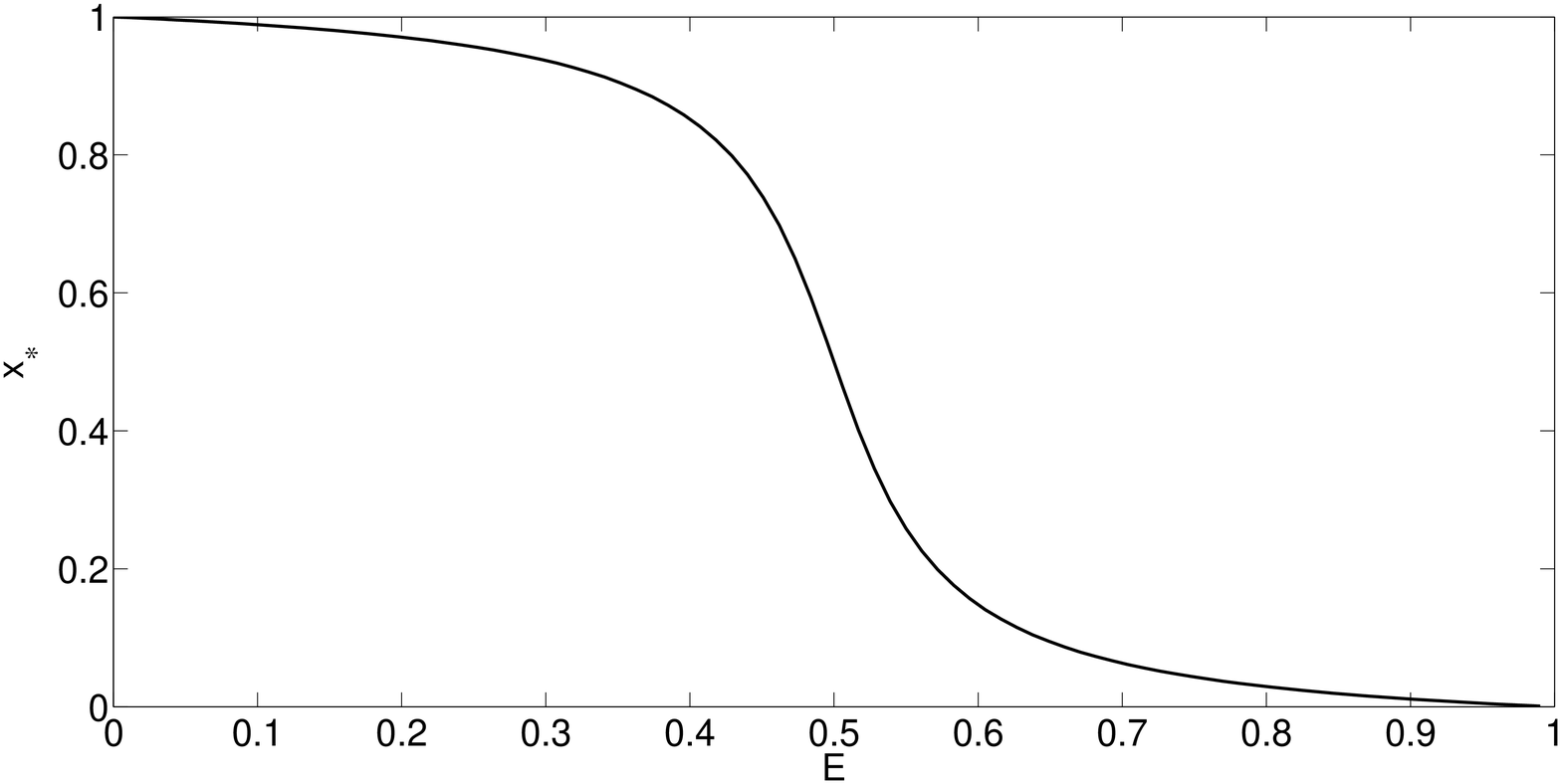}
\caption{
Plot of $x_\ast(E)$ solution (\ref{equilibrioES}), $K_M = 0.1$. 
}
\label{stab}
\end{figure}

\begin{figure}
\includegraphics[width=0.5\textwidth]
{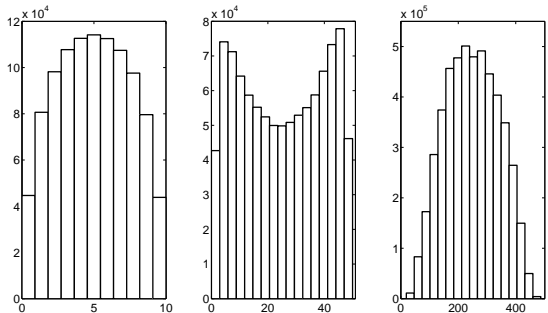}
\caption{
Histograms of $x$ for $N=10, 50, 500$, obtained by numerical simulations run for $5\cdot10^6$ iterations. The ratio between $M$ (enzyme molecule number) and $N$ (substrate molecule number) is set to $0.15$; the constants for enzyme production reactions $k_1$, $k_{-1}$ are set to $2$, $K_C=100$, $K_M=0.1$.
}
\label{SizeMatters}
\end{figure}

In Fig. \ref{stab} we show the plot of $x_\ast(E)$ with $K_M = 0.1$ to point out the sigmoidal behaviour due to the extreme sensitivity of the solution to enzyme concentration.
According to (\ref{p_s_s}) the substrate distribution $\overline{p}_s(x)$ can be computed:
\begin{equation}
\overline{p}_s(x)\propto \int_0^1 \exp\left ( -\frac{|C'(x_\ast)|}{D(x_\ast)}\frac{(x-x_\ast)^2}{2} \right ) p(x_\ast)d x_\ast
\label{f_dist}
\end{equation}
where $ p(x_\ast)$ is the critical point distribution defined by (\ref{prob_dist}) and we use the Gaussian approximation for the FP stationary distribution. If $p(x_\ast)$ is a bimodal distribution (i.e. the condition (\ref{crit_cond}) is satisfied), the final distribution $\overline{p}_s(x)$ will also be bimodal when the variance $D(x_\ast)/|C'(x_\ast)|$ is sufficiently small: indeed smaller than the distance of the critical points of $p(x_\ast)$. 
In general, this condition is satisfied if $(K_M N)^{-1}$ is sufficiently small, which is consistent with the choice $N\gg 1$.
In Figure \ref{SizeMatters}, we perform numerical simulations of the CME, using a low number of enzyme and substrate molecules. We observe that bimodality occurs only under suitable conditions depending on the system size: fixing the chemical reaction constants and the ratio between enzyme and substrate total molecule number $E_T/X_T$, only in a specific interval of molecule number (say $X_T$, the total number of substrate molecules) we have such behaviour. The condition (\ref{crit_cond}) can be applied only in the last case ($N=500$) and it is consistent with numerical simulations (loss of bimodality).

This phenomenon can be explained as follows: increasing the molecule number, we recover the Michaelis-Menten deterministic limit, losing the noise-induced bistability; on the other side, if the molecule number is too low, even if bistability is not lost from an analytical point of view, the separation between the two maxima becomes indistinguishable.

\section{Conclusions}
In this paper we study the conditions that result in a bimodal stationary distribution of a single phosphorylation/dephosphorylation cycle described by Michaelis-Menten equations in the quasi-steady state assumption.
The simple addition of noise (as produced by a Master Equation approach) is not enough to achieve bimodality in this system that has a unique deterministic stable state, and the same result can be stated for the Fokker-Planck description of the system obtained as a limit of the Master Equation. Since the Fokker-Planck approach is often used instead of the exact Master Equation approach, we characterize the conditions in which it approximates correctly the system.

We propose a different approach to generate stochastic bimodality (that does not change the number of deterministic stable states, thus a purely noise-driven phenomenon). With an additional chemical reaction we introduce competition between enzymes, for binding to an activating molecule or for reaching a specific site in which reaction can occur (e.g. an organelle), that can be biologically feasible. 
For this system we clarify the conditions for which stochastic fluctuations in enzyme concentration can lead to bimodality in substrate concentration, and we show that it depends on the time scales involved in the two reactions (related to the kinetic constants) and also on the size of the system (i.e. the number of enzyme and substrate molecules involved). 

In particular, if we fix the kinetic constants and the ratio between enzyme and substrate molecule numbers, bimodality is observed only in specific parameter s, related to system size in terms of total number of molecules involved. This results define in more detail the feasibility of this phenomenon in real biological systems, stating that if the biological systems have to exploit noise to achieve a bimodal behaviour there must be a relationship between the chemical parameters of the system and its size.

\section{Appendix}
The FP approximation (\ref{FP_staz}) for the stationary solution (\ref{boltz}) requires that the condition $D_+V(n)=-\ln g_n/r_{n+1}$
reduces to
$$
\frac{dV}{dx}=-2\frac{C(x)}{D(x)}+\frac{\partial}{\partial x}\ln D(x)
$$
where in the thermodynamics limit $N\to\infty$ where $x=n/N$ and $r_{n+1/2}=Nr(x)$, $g_{n+1/2}=Ng(x)$.
This is the case when $g_n-r_{n+1}$ is small so that using a perturbative
expansion we have
\begin{eqnarray*}
\ln \frac{g_n}{r_{n+1}}\simeq 2\frac{r_{n+1}-g_n}{r_{n+1}+g_n} \simeq\\
\frac{1}{N}\left [\frac{N(r(x)-g(x))+\frac{1}{2}\frac{\partial}{\partial x}(r(x)+g(x))}{r(x)+g(x)}  \right ]+O\left (\frac{[r(x)-g(x)]^2}{N}\right )
\end{eqnarray*}
Therefore if $N(r(x)-g(x))$ is finite we can approximate
$$
\frac{dV}{dx}\simeq ND_+V(n)\simeq \left [\frac{N(r(x)-g(x))+\frac{1}{2}\frac{\partial}{\partial x}(r(x)+g(x))}{r(x)+g(x)}  \right ]
$$
with an error of order $O([r(x)-g(x)]^2)$. Finally we get
\begin{eqnarray*}
\frac{d V}{dx}\simeq -2N\frac{g(x)-r(x)}{r(x)+g(x)}+\frac{\partial}{\partial x}\ln (r(x)+g(x)) \\
= -2\frac{C(x)}{D(x)}+\frac{\partial}{\partial x}\ln D(x)
\end{eqnarray*}
The condition $g_n-r_{n+1}\ll 1$ (i.e. the generation and recombination
rates of the enzymatic cycle have almost the same value) means that the FP approximation is justified only nearby the critical points when
$g_n/r_{n+1}\simeq 1$.


\bibliography{CMEarkin} 

\end{document}